\documentclass[pra,showpacs,twocolumn,amsmath,amssymb]{revtex4}
\usepackage{graphicx}
\begin{document}
\title{Linear optics substituting scheme for multi-mode operations}
\date{\today}
\author{J. Clausen}
\email{J.Clausen@tpi.uni-jena.de}
\author{L. Kn\"oll}
\author{D.-G. Welsch}
\affiliation{Friedrich-Schiller-Universit\"at Jena,\\
             Theoretisch-Physikalisches Institut,\\
             Max-Wien-Platz 1, D-07743 Jena, Germany}
\begin{abstract}
We propose a scheme allowing a conditional implementation of suitably truncated
general single- or multi-mode operators acting on states of traveling optical
signal modes. The scheme solely relies on single-photon and coherent states and
applies beam splitters and zero- and single-photon detections. The signal flow
of the setup resembles that of a multi-mode quantum teleportation scheme thus
allowing the individual signal modes to be spatially separated from each other.
Some examples such as the realization of cross-Kerr nonlinearities, multi-mode
mirrors, and the preparation of multi-photon entangled states are considered.
\end{abstract}
\maketitle
\section{\label{sec1}
         Introduction}
A general problem in quantum optics is the implementation of a defined
single- or multi-mode operator by some physical scheme since many efforts
amount to the realization of a desired quantum state transformation of an
unknown or control of a known signal quantum state
\cite{lloyd1,lloyd2,lloyd3,lloyd4,lloyd5,lloyd6,
akulin1,akulin2,akulin3,akulin4,
schirmer3,schirmer4,schirmer5,
drobny1,drobny2,drobny3}.
Here, we limit attention to a physical system consisting of a number of
traveling light pulses described by non-monochromatic signal modes. We assume
that the spatial pulse length is large compared to the wavelength of the
radiation, so that we deal with quasi-monochromatic pulses corresponding to
quasi-orthogonal modes.

A difficulty is that at present, only a limited set of basic operations can be
implemented directly. Composite operations therefore have to be constructed from
elementary ones which are simple enough to allow their direct realization.
Examples of such basic operations are the preparation of coherent (i.e.,
Glauber) states by single-mode lasers, or single-photon states by parametric
down converters, furthermore parametric interactions as realized by beam
splitters, three-wave mixers, and Kerr-nonlinearities, and the discrimination
between presence and absence of photons by means of binary
{{0}}-{{1}}-photodetectors such as Avalanche photodiodes. A simple combination
of these techniques allows further manipulations such as parametric
amplification, coherent displacement, or the preparation and detection of
photon number (i.e., Fock) states \cite{braunstein1}.

To give an example, the signal pulse may be sent through a medium applied in the
parametric approximation, i.e., the medium realizes a weak coupling between the
signal modes and a number of auxiliary modes prepared in strong coherent states.
Since within classical optics, $n$th-order interactions are described as a
$n$th-order deviation from linearity of the polarization induced in a medium by
an electric field, the interaction strength is expected to decline rapidly with
increasing order. Strong fields are therefore required for their observation.
In contrast, setups discussed within quantum optics and quantum information
processing often operate with superpositions of low-excited Fock states while
the coherent amplitudes in the auxiliary modes cannot be increased unlimited to
achieve a desired interaction strength of the reduced signal operation. This
greatly limits the order of the nonlinearity applicable and with it the variety
of unitary transformations that can be realized by a given medium. Desirable are
therefore substituting schemes for such nonlinear interactions
\cite{fiurasek1,vogel1}. In particular, one may apply the nonlinearity hidden in
the quantum measurement process and to use merely passive optical elements such
as beam splitters \cite{zeilinger1} and photodetectors while relying on simple
auxiliary preparations such as coherent and single-photon states
\cite{linearMilburn}. A local measurement performed on a spatially extended
quantum system affects the reduced state at the other locations. By repeated
measurements and postselection of desired detection events, this back-action
caused by a measurement can then be used to perform well-defined manipulations,
including state preparations
\cite{steuernagel1,paris1,koashi2,zou2,zou3,zou4,cerf1,fiurasek3},
quantum logic operations \cite{koashi1,ralph1,pittman1,zou1,ralph2,zou5},
state purifications \cite{zeilinger2}, quantum error corrections
\cite{gottesman1}, and state detections \cite{dusek1,calsamiglia2}. The simplest
basic operations one may think of are the application of the mode operators
$\hat{a}$ and $\hat{a}^\dagger$, i.e., photon subtraction and addition
\cite{dakna1,calsamiglia1}. With regard to compositions, it is necessary to
limit operation to a suitably chosen finite-dimensional subspace of the
system's Hilbert space, cf., e.g., \cite{gottesman1}, since only a finite number
of parameters can be controlled.

The aim of this paper is to investigate a theoretical possibility of
implementing a desired single- or multi-mode operation $\hat{Y}$ on the quantum
state $\hat{\varrho}$ of a traveling optical signal. The scheme solely relies on
beam splitters as well as zero- and single-photon detections. In this way,
nonlinear and active optical elements can be avoided. It requires the
preparation of coherent states and single-photon states, however. The idea is to
start with a single photon, which is then manipulated to construct an entangled
$k$-mode state. After that, the latter is shared by local setups performing the
transformation in the signal modes. The measurement-assisted and hence
conditional transformation leads to an output state
\begin{equation}
\label{Y}
  \hat{\varrho}^\prime=\frac{1}{p}\hat{Y}\hat{\varrho}\hat{Y}^\dagger,
\end{equation}
where the normalization constant
\begin{equation}
\label{p}
  p=\mathrm{Tr}(\hat{Y}\hat{\varrho}\hat{Y}^\dagger)
  =\bigl\langle\hat{Y}^\dagger\hat{Y}\bigr\rangle
\end{equation}
is the probability of the respective measurement result, i.e. the
\textquoteleft{success probability}\textquoteright\,. In particular, if the
operator $\hat{Y}$ is proportional to a unitary one,
$\hat{Y}$ $\!=$ $\!\alpha\hat{U}$, where
$\hat{U}^\dagger$ $\!=$ $\!\hat{U}^{-1}$, then the success probability is
independent of the input state, $p$ $\!=$ $\!|\alpha|^2$.

Within the scope of explaining the principle, we limit attention to idealized
optical devices, as the effect of imperfections depends on the desired
transformation and the signal state itself. In a given practical setup, the
imperfections and the resulting coupling of the system to its environment must
be considered since loss of only one photon may change the phase in a
superposition of two states and yield wrong results. Apart from this, the mode
matching becomes an important issue especially in the case of composite devices
such as optical multiports. A generalization of Eq.~(\ref{Y}) allowing the
inclusion of loss would be the transformation $\hat{\varrho}^\prime$ $\!=$
$\!p^{-1}\sum_l\kappa_l\hat{Y}_l\hat{\varrho}\hat{Y}_l^\dagger$ with given
coefficients $\kappa_l$ $\!\ge$ $\!0$.

The article is organized as follows. Section~\ref{sec2} describes the local
devices which mix the signal pulses and an entangled state in order to perform
the desired transformation of the signal, whereas section~\ref{sec3} is
dedicated to the preparation of the entangled state itself. The operation of the
complete setup is considered in section~\ref{sec4}, and section~\ref{sec5}
explains how general operators can be approached with it. To give some examples,
a few operators of special interest are considered in section~\ref{sec6},
such as functions of the single-mode photon number operator in
section~\ref{sec6.1}, two-mode cross-Kerr interactions in section~\ref{sec6.2},
U($N$)-transformations in section~\ref{sec6.3}, and the preparation of
multi-photon entangled states in section~\ref{sec6.4}. Finally, a summary and
some concluding remarks are given in section~\ref{sec7}. An appendix is added
to outline three possibilities of implementing single-photon cloning as required
to prepare the entangled state and some ordering relations for the photon number
operator.
\section{\label{sec2}
         Local operations}
\begin{figure}[ht]
\includegraphics[width=8.6cm]{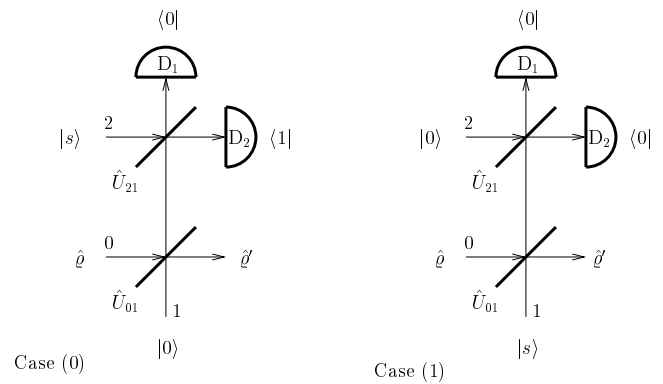}
\caption{\label{fig1}
Setup consisting of beam splitters $\hat{U}$ and photodetectors D and
implementing conditional subtraction [case (0)] or addition [case (1)] of a
photon on a signal state $\hat{\varrho}$, each controlled by the photon number
$s$ $\!=$ $\!0,1$ of the Fock state $|s\rangle$, respectively. Note that in case
(1) the beam splitter $\hat{U}_{21}$ is redundant.
}
\end{figure}

The scheme consists of $k$ local devices, each implementing a conditional
single-mode photon subtraction or addition, depending on the case in which it is
applied. The setup of such a local device is shown in Fig.~\ref{fig1} and
consists of beam splitters $\hat{U}_{01}(T_1,R_1)$ and $\hat{U}_{21}(T_2,R_2)$
as well as photodetectors D$_1$ and D$_2$. We describe the mixing of some mode
$j$ with some other mode $k$ by a beam splitter in the usual way by a unitary
operator $\hat{U}_{jk}(T,R,P)$ defined via the relation
\begin{equation}
\label{MC}
  \hat{U}_{jk}^{\dagger}\binom{\hat{a}_j}{\hat{a}_k}\hat{U}_{jk}^{}
  =P\left(\begin{array}{cc}
  T & R \\
  -R^* & T^*
  \end{array}\right)
  \binom{\hat{a}_j}{\hat{a}_k}
\end{equation}
by its complex transmittance $T$, reflectance $R$, and phase $P$ obeying
$|T|^2$ $\!+$ $\!|R|^2$ $\!=$ $\!|P|^2$ $\!=$ $\!1$. If not stated otherwise,
we will assume that $P$ $\!=$ $\!1$ within the work.

The setup shown in Fig.~\ref{fig1} realizes a transformation of the signal state
$\hat{\varrho}$ in mode 0 according to Eq.~(\ref{Y}), where the single-mode
operator $\hat{Y}$ $\!=$ $\!\hat{Y}_0(s)$ depends on the photon number
$s$ $\!=$ $\!0,1$ of the Fock state $|s\rangle$ in which one of the remaining
input modes are prepared. We distinguish between two different cases of
operation.
First consider case (0), in which input mode 1 is prepared in the vacuum state
$|0\rangle$ and input mode 2 in $|s\rangle$. If D$_1$ and D$_2$ detect 0 and 1
photons, respectively, the operator $\hat{Y}_0(s)$ becomes
\begin{eqnarray}
  \hat{Y}_0^{(0)}(s)
  &=&\,_1\langle0|\,_2\langle1|\hat{U}_{21}\hat{U}_{01}|0\rangle_1|s\rangle_2
  \nonumber\\
\label{Y0}
  &=&R_2\left(\frac{T_2}{R_2}\right)^s
  T_1^{\hat{n}_0}(-R_1^*\hat{a}_0)^{1-s},
\end{eqnarray}
where $\hat{n}_j$ $\!=$ $\!\hat{a}_j^\dagger\hat{a}_j$ is the photon number
operator. In case (1), input mode 2 is prepared in the vacuum state $|0\rangle$
and input mode 1 in $|s\rangle$. If D$_1$ and D$_2$ detect 0 photons each, the
operator $\hat{Y}_0(s)$ reads
\begin{eqnarray}
  \hat{Y}_0^{(1)}(s)
  &=&\,_1\langle0|\,_2\langle0|\hat{U}_{21}\hat{U}_{01}|s\rangle_1|0\rangle_2
  \nonumber\\
\label{Y1}
  &=&T_1^{\hat{n}_0}\left(\frac{R_1}{T_1}\hat{a}_0^\dagger\right)^s.
\end{eqnarray}
We see that (apart from the operator $T_1^{\hat{n}_0}$ which is always present)
in case (0) we realize a photon subtraction and in case (1) a photon addition,
controlled by the photon number $s$ $\!=$ $\!0,1$ of the control input state
$|s\rangle$.
\section{\label{sec3}
         Preparation of the entangled state}
The combined operation of $k$ local setups shown in Fig.~\ref{fig1} requires an
entangled $k$-mode state
\begin{equation}
\label{sk}
  |\Psi\rangle_{1,\ldots,k}
  =\frac{|s_1\rangle_1\cdots|s_k\rangle_k
  +z|1-s_1\rangle_1\cdots|1-s_k\rangle_k}{\sqrt{1+|z|^2}}
\end{equation}
shared by all the stations. Here, $z$ is a given complex number, and the
$|s\rangle$ are photon number states with $s$ $\!=$ $\!0,1$. In this section, we
discuss the preparation of the state Eq.~(\ref{sk}). We start with preparing a
mode 0 in a single-photon state $|1\rangle$ and mix it with some other mode $j$
prepared in the vacuum state $|0\rangle$ using a balanced beam splitter
$\hat{U}_{0j}(T$ $\!=$ $\!-R$ $\!=$ $\!1/\sqrt{2})$, which leaves the photon in
a state
\begin{equation}
\label{s}
  |\Psi\rangle_{0j}
  =\hat{U}_{0j}|1\rangle_0|0\rangle_j
  =\frac{|0\rangle_0|1\rangle_j+|1\rangle_0|0\rangle_j}{\sqrt{2}}.
\end{equation}
By a repeated application of single-photon cloners
\begin{equation}
\label{Q}
  \hat{Q}_{kj}=\sum_{s=0}^1|s\rangle_k|s\rangle_j\,_j\langle s|,
\end{equation}
which duplicate photon number states $|s\rangle$ with $s$ $\!=$ $\!0,1$
according to $\hat{Q}_{kj}|s\rangle_j$ $\!=$ $\!|s\rangle_k|s\rangle_j$, the
state Eq.~(\ref{s}) can now be enlarged to a ($k$ $\!+$ $\!1$)-mode state
\begin{eqnarray}
  |\Psi\rangle_{0,\ldots,k}
  &=&\hat{Q}_{kj}\cdots\hat{Q}_{j+1,j}\hat{Q}_{j-1,0}\cdots\hat{Q}_{10}
  |\Psi\rangle_{0j}
  \nonumber\\
\label{sk+1}
  &=&\frac{|0\rangle_{0,\ldots,j-1}|1\rangle_{j,\ldots,k}
  +|1\rangle_{0,\ldots,j-1}|0\rangle_{j,\ldots,k}}{\sqrt{2}},\;\;\;
\end{eqnarray}
where $0$ $\!<$ $\!j$ $\!\le$ $\!k$ and $k$ $\!>$ $\!0$. Note that here we have
used the notation
$|s\rangle_{m,\ldots,n}$ $\!=$ $\!|s\rangle_m|s\rangle_{m+1}\cdots|s\rangle_n$.
Some possibilities allowing the implementation of Eq.~(\ref{Q}) are outlined in
appendix~\ref{secA1}, cf. also \cite{fiurasek2}. To manipulate the state
Eq.~(\ref{sk+1}) further, we mix mode 0 with an auxiliary mode $k$ $\!+$ $\!1$
prepared in a coherent state
$|\alpha\rangle$ $\!=$ $\!\mathrm{e}^{-\frac{|\alpha|^2}{2}}
\sum_{n=0}^\infty\alpha^n(n!)^{-\frac{1}{2}}|n\rangle$
using a beam splitter $\hat{U}_{0,k+1}(T,R)$. If thereafter 1 and 0 photons are
detected in mode 0 and $k$ $\!+$ $\!1$, respectively, the state Eq.~(\ref{sk+1})
is reduced to the $k$-mode state Eq.~(\ref{sk}),
\begin{equation}
  \,_0\langle1|\,_{k+1}\langle0|\hat{U}_{0,k+1}|\Psi\rangle_{0,\ldots,k}
  |\alpha\rangle_{k+1}
  =\mathrm{e}^{\mathrm{i}\varphi}\sqrt{p}\;|\Psi\rangle_{1,\ldots,k}.
\end{equation}
Here, $\varphi$ $\!=$ $\!\arg{T}$ is an unrelevant phase and the photon numbers
$s_1,$ $\!\ldots,$ $\!s_k$ can be chosen arbitrarily to be either 0 or 1 by
permutating the mode indices suitably, except that one of the $s_l$ is always 0.
Note that the latter does not constitute a limitation since we have
\begin{equation}
  |\Psi(s_1,\cdots,s_k;z)\rangle
  =\mathrm{e}^{\mathrm{i}\arg{z}}
  |\Psi(1\!-\!s_1,\cdots,1\!-\!s_k;z^{-1})\rangle.
\end{equation}
The complex parameter $z$ $\!=$ $\!T^{-1}R\alpha$ can be controlled by $T$ or
$\alpha$. It is however convenient to maximize the success probability
\begin{equation}
  p=\frac{(1+|z|^2)|\alpha|^2\mathrm{e}^{-|\alpha|^2}}{2(|\alpha|^2+|z|^2)}
\end{equation}
by adjusting $|T^{-1}R|$ such that
\begin{equation}
  |\alpha|^2=\sqrt{\frac{|z|^4}{4}+|z|^2}-\frac{|z|^2}{2}
\end{equation}
is valid, in which case it is ensured that
$(2\mathrm{e})^{-1}$ $\!<$ $\!p$ $\!<$ $\!2^{-1}$ holds for all $z$.
\section{\label{sec4}
         Overall operation}
\begin{figure}[ht]
\includegraphics[width=8.6cm]{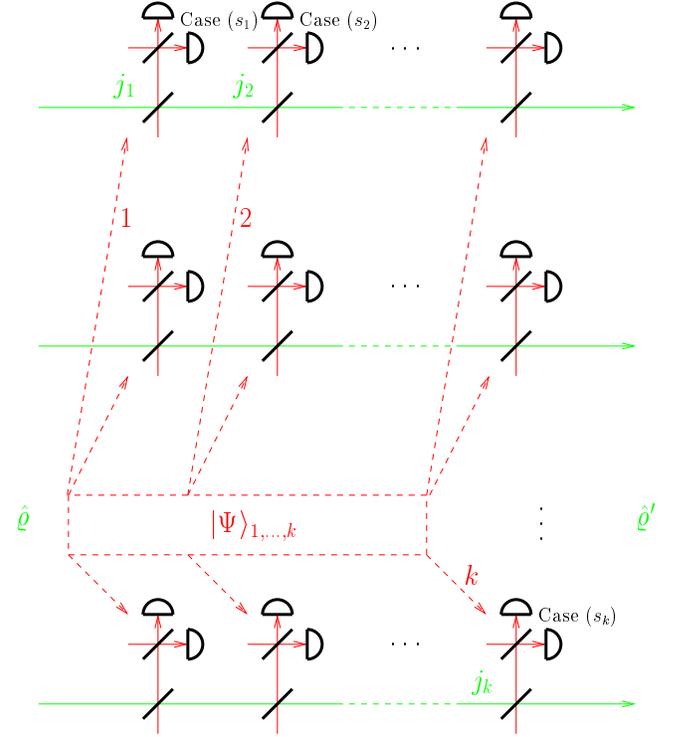}
\caption{\label{fig2}
Example of the overall setup consisting of $k$ devices Fig.~\ref{fig1} applied
in the signal modes $j_1,\ldots,j_k$ in cases $(s_1),\ldots,(s_k)$,
respectively. These modes do not have to be distinct from another. The
respective control input ports of the devices which are fed with states
$|s\rangle$ in Fig.~\ref{fig1}, here share an entangled state
$|\Psi\rangle_{1,\ldots,k}$ given by Eq.~(\ref{sk}).
}
\end{figure}

Let us now assume that a state $|\Psi\rangle_{1,\ldots,k}$ given by
Eq.~(\ref{sk}) is prepared. Each mode $l$ $\!=$ $\!1,\ldots,k$ is fed into the
control input of a device Fig.~\ref{fig1} (the input port fed with the variable
photon number state $|s\rangle$ in Fig.~\ref{fig1}), assuming case $(s_l)$,
respectively. A schematic view of an example of the overall setup is depicted in
Fig.~\ref{fig2}.
Denoting the respective signal mode (mode 0 in Fig.~\ref{fig1}) by
$j_l$, we obtain a transformation in modes $j_1,$ $\!\ldots,$ $\!j_k$ of the
signal state $\hat{\varrho}$ according to Eq.~(\ref{Y}), where
\begin{eqnarray}
  \hat{Y}(\hat{A},\alpha,\beta)
  &=&\frac{1}{\sqrt{1+|z|^2}}
  \Bigl[\hat{Y}_{j_k}^{(s_k)}(s_k)\cdots\hat{Y}_{j_1}^{(s_1)}(s_1)
  \nonumber\\
  &&+z\hat{Y}_{j_k}^{(s_k)}(1-s_k)\cdots\hat{Y}_{j_1}^{(s_1)}(1-s_1)\Bigr]
  \nonumber\\
\label{Yk}
  &=&\alpha\hat{T}(\hat{A}+\beta),
\end{eqnarray}
with
\begin{subequations}
\label{Ykpar-a}
\begin{eqnarray}
  \hat{T}&=&T_1^{\sum_{l=1}^k\hat{n}_{j_l}},
  \\
\label{Ykparb}
  \hat{A}&=&\hat{a}_{j_k}^{(s_k)}\cdots\hat{a}_{j_1}^{(s_1)},
  \\
  \alpha&=&\frac{z}{\beta\sqrt{1+|z|^2}}T_2^{k-\sum_{l=1}^ks_l},
  \\
  \beta&=&zR_1^{-\sum_{l=1}^ks_l}
  \left(-\frac{T_1^{}}{R_1^*}\frac{T_2^{}}{R_2^{}}\right)^{k-\sum_{l=1}^ks_l}
  \nonumber\\
  &&\times T_1^{\sum_{r=1}^k\sum_{l=r}^k(2s_l-1)\delta_{j_r,j_l}},
\end{eqnarray}
\end{subequations}
and $\hat{a}^{(s)}$ stands for $\hat{a}$ if $s$ $\!=$ $\!0$ and for
$\hat{a}^\dagger$ if $s$ $\!=$ $\!1$. We see that for given beam splitter
parameters $T_k$ and $R_k$, $\beta$ can be chosen as desired by varying $z$.
Note that the signal modes $j_1,$ $\!\ldots,$ $\!j_k$ don't have to be distinct,
cf. the examples in section~\ref{sec6.1} and section~\ref{sec6.2}. Note also
that those modes which are distinct may be spatially separated from each other.
The signal flow of the scheme Fig.~\ref{fig2} then resembles that of a
multi-mode quantum teleportation setup.

We now consider $N$ successive applications of Eq.~(\ref{Yk}). If $z$ is varied
from step to step, the resulting overall operator becomes the product of the
individual operators Eq.~(\ref{Yk}) according to
\begin{eqnarray}
  \hat{Y}_\mathrm{tot}&=&
  \hat{Y}(\hat{A},\alpha_N,\beta_N)\cdots\hat{Y}(\hat{A},\alpha_1,\beta_1)
  \nonumber\\
  &=&\hat{\prod}_{n=1}^N\left[\alpha_n\hat{T}(\hat{A}+\beta_n)\right]
  \nonumber\\
  &=&\gamma^{\frac{N(N-1)}{2}}\hat{T}^N\prod_{n=1}^N\left[\alpha_n(\hat{A}
  +\beta_n\gamma^{1-n})\right]
  \nonumber\\
\label{Y+-N}
  &=&\gamma^{\frac{N(N-1)}{2}}\left(\prod_{n=1}^N\alpha_n\right)
  f_N^{-1}\hat{T}^NF_N(\hat{A}),
\end{eqnarray}
where $\alpha_n$ and $\beta_n$ follow according to Eqs.~(\ref{Ykpar-a}) from the
parameters of $\hat{U}_{01}$ and $\hat{U}_{21}$ during the $n$'th passage and
$\gamma$ is given by $\hat{A}\hat{T}$ $\!=$ $\!\gamma\hat{T}\hat{A}$, so that
\begin{equation}
\label{gamma}
  \gamma=T_1^{\sum_{r=1}^k\sum_{l=1}^k(1-2s_l)\delta_{j_r,j_l}},
\end{equation}
and
\begin{equation}
\label{poly}
  F_N(\hat{A})
  =\sum_{n=0}^Nf_n\hat{A}^n
  =f_N\prod_{n=1}^N(\hat{A}+\beta_n\gamma^{1-n})
\end{equation}
is an arbitrary polynomial of order $N$ in $\hat{A}$ determined by its roots
$-\beta_n\gamma^{1-n}$. The product symbol $\hat{\prod}$ has been used to
indicate that the index of the factors increases from right to left.

Assume now that we want to implement an arbitrary expandable function
$F(\hat{A})$. In general, $\hat{A}$ can only be approximated by a polynomial
$F_N(\hat{A})$ of order $N$ in $\hat{A}$ with sufficiently large $N$. There are
however two cases in which $F(\hat{A})$ can be replaced exactly with some
polynomial.
\subsection{\label{sec4.1}
            Existence of an eigenspace of $\hat{A}$}
In the first case, a set of $N^\prime$ $\!+$ $\!1$ eigenstates
$|\varphi_{l^\prime}\rangle$ of $\hat{A}$ can be found,
\begin{equation}
\label{eigenstates}
  \hat{A}|\varphi_{l^\prime}\rangle
  =A^\prime_{l^\prime}|\varphi_{l^\prime}\rangle
  =A_l^{}|\varphi_{l^\prime}\rangle,
  \quad
  \substack{l=0,\ldots,N\;\;,\\l^\prime=0,\ldots,N^\prime,}
\end{equation}
where $N^\prime$ ($\!\ge$ $\!N$) is chosen such that a renumbering of the
eigenvalues $A_{l^\prime}^\prime$ gives a set $\{A_0,\ldots,A_N\}$ of exactly
$N$ $\!+$ $\!1$ distinct $A_l$. We now choose Eq.~(\ref{poly}) to be the
polynomial
\begin{equation}
\label{eigenpoly}
  F_N(\hat{A})
  =\sum_{l=0}^N
  F(A_l)\prod_{\substack{k=0\\(k\neq l)}}^N\frac{\hat{A}-A_k}{A_l-A_k}
\end{equation}
of order $N$ in $\hat{A}$, so that we obtain
$F_N(\hat{A})|\varphi_{l^\prime}\rangle$
$\!=$ $\!F(\hat{A})|\varphi_{l^\prime}\rangle$. In Eq.~(\ref{Y+-N}) we then have
\begin{equation}
\label{Ytot}
  \hat{Y}_\mathrm{tot}\hat{P}_{N^\prime}^{(j_1,\ldots,j_k)}
  =\frac{\prod_{n=1}^N\alpha_n}{f_N}\gamma^{\frac{N(N-1)}{2}}
   \hat{T}^NF(\hat{A})\hat{P}_{N^\prime}^{(j_1,\ldots,j_k)},
\end{equation}
where
\begin{equation}
  \hat{P}_{N^\prime}^{(j_1,\ldots,j_k)}
  =\sum_{l^\prime=0}^{N^\prime}
  |\varphi_{l^\prime}\rangle\langle\varphi_{l^\prime}|.
\end{equation}
We see that a signal state $\hat{\varrho}$ satisfying the condition
\begin{equation}
  \hat{\varrho}\hat{P}_{N^\prime}^{(j_1,\ldots,j_k)}
  =\hat{P}_{N^\prime}^{(j_1,\ldots,j_k)}\hat{\varrho}
  =\hat{\varrho}
\end{equation}
is transformed according to Eq.~(\ref{Y}), where
$\hat{Y}$ $\!\sim$ $\!\hat{T}^NF(\hat{A})$, cf. Eq.~(\ref{Ytot}). In this
sense, a desired function $F(\hat{A})$ can be implemented. In order to
compensate the operator $\hat{T}$, the transmittance $T_1$ may be chosen
sufficiently close to unity, i.e., $|R_1|$ $\!\ll$ $\!1$. A precise compensation
is possible by an additional implementation of the operator $\hat{T}^{-N}$
allowed in the photon-number truncated case as discussed below, in
section~\ref{sec6.1}. We then have $\hat{Y}$ $\!\sim$ $\!F(\hat{A})$.
Note that the replacement of the function $F(\hat{A})$ by a polynomial
$F_N(\hat{A})$ of order $N$ in $\hat{A}$ is unique. If there was another
polynomial $G_N(\hat{A})$ of order $N$ in $\hat{A}$ for which also
$G_N(\hat{A})|\varphi_{l^\prime}\rangle$
$\!=$ $\!F(\hat{A})|\varphi_{l^\prime}\rangle$ with
$l^\prime$ $\!=$ $\!0,$ $\!\ldots,$ $\!N^\prime$, we could consider the
polynomial $F_N(x)$ $\!-$ $\!G_N(x)$ of order $N$ in $x$. It has however the
($N$ $\!+$ $\!1$) distinct roots $x_l$ $\!=$ $\!A_l$, so that
$F_N(x)$ $\!-$ $\!G_N(x)$ $\!\equiv$ $\!0$.
Of particular interest is the case in which in each of the modes
$j_1,$ $\!\ldots,$ $\!j_k$, the number of annihilation operators occuring in
$\hat{A}$ equals the number of creation operators, so that the eigenstates
$|\varphi_{l^\prime}\rangle$ $\!=$
$\!|n_{j_1}^{(l^\prime)},\ldots,n_{j_k}^{(l^\prime)}\rangle$ are
$N^\prime$ $\!+$ $\!1$ photon number states (they may be, e.g., the lowest
states occuring in a Fock space expansion).
\subsection{\label{sec4.2}
            Photon-number truncated signal states}
In the other case, there exists a mode $j_l$ in which the number of annihilation
operators occuring in $\hat{A}$ exceeds the number of creation operators and the
signal state $\hat{\varrho}$ is photon-number truncated in this mode,
\begin{equation}
\label{truncsig}
  \hat{\varrho}\hat{P}_N^{(j_l)}
  =\hat{P}_N^{(j_l)}\hat{\varrho}
  =\hat{\varrho},
  \quad\quad\hat{P}_N^{(j_l)}
  =\sum_{n=0}^N|n\rangle_{j_l}\,_{j_l}\langle n|.
\end{equation}
We choose Eq.~(\ref{poly}) to be the sum of the first $N$ $\!+$ $\!1$ elements
of the power series expansion of $F(\hat{A})$, so that the signal state
$\hat{\varrho}$ is again transformed according to Eq.~(\ref{Y}), where
$\hat{Y}$ $\!\sim$ $\!\hat{T}^NF(\hat{A})$, cf. Eq.~(\ref{Ytot}).
\section{\label{sec5}
         Approaching general operator functions}
Assume that we want to implement an arbitrary function $\hat{F}$ of the creation
and annihilation operators of a number of signal modes,
$\hat{Y}$ $\!\sim$ $\!\hat{F}$ in Eq.~(\ref{Y}). We further assume that the
expansion of $\ln\hat{F}$ can according to
\begin{equation}
 \hat{F}=\mathrm{e}^{\ln\hat{F}}=\mathrm{e}^{\sum_{n=1}^\infty
  c_n\hat{\mathcal{A}}_n}
  \approx\mathrm{e}^{\sum_{n=1}^Nc_n\hat{\mathcal{A}}_n}
\end{equation}
be truncated after a (sufficiently large) number $N$ of elements. Here, the
$\hat{\mathcal{A}}_n$ are products of the respective creation and annihilation
operators that can be written in the form of Eq.~(\ref{Ykparb}). Consider now
$N^2$ successive applications of Eq.~(\ref{Yk}) according to
\begin{eqnarray}
  \hat{Y}_\mathrm{tot}&=&
  \hat{Y}(\hat{A}_{N^2},\alpha_{N^2},\beta_{N^2})
  \cdots\hat{Y}(\hat{A}_1,\alpha_1,\beta_1)
  \nonumber\\
  &=&\hat{\prod}_{n=1}^{N^2}
  \left[\alpha_n\hat{T}(\hat{A}_n+\beta_n)\right]
  \nonumber\\
  &=&\hat{T}^{N^2}\hat{\prod}_{n=1}^{N^2}
  \left[\alpha_n\beta_n(1+\beta_n^{-1}\gamma_n^{n-1}\hat{A}_n)\right],
\end{eqnarray}
where $\gamma_n$ is given by $\hat{A}_n\hat{T}$ $\!=$
$\!\gamma_n\hat{T}\hat{A}_n$, cf. Eq.~(\ref{gamma}). Choosing $\hat{A}_n$ as
well as $\beta_n$ according to
\begin{subequations}
\begin{eqnarray}
  \hat{A}_n&=&\hat{\mathcal{A}}_{\lceil{n}\rceil},
  \\
  \beta_n&=&Nc_{\lceil{n}\rceil}^{-1}\gamma_n^{n-1},
\end{eqnarray}
\end{subequations}
where $\lceil{n}\rceil$ $\!\equiv$
$\!1$ $\!+$ $\![(n$ $\!-$ $\!1)$ $\!\mathrm{mod}$ $\!N]$, we see that
\begin{equation}
\label{YgenN2}
  \hat{Y}_\mathrm{tot}
  =\biggl(\prod_{n=1}^{N^2}\alpha_n\beta_n\biggr)\hat{T}^{N^2}\hat{X},
\end{equation}
where
\begin{equation}
  \hat{X}=
  \left[\hat{\prod}_{n=1}^N\left(1+N^{-1}c_n\hat{\mathcal{A}}_n\right)\right]^N
  \stackrel{(N\gg1)}{\approx}
  \mathrm{e}^{\sum_{n=1}^Nc_n\hat{\mathcal{A}}_n},
\end{equation}
and therefore $\hat{X}$ $\!\approx$ $\!\hat{F}$ for $N\gg1$. Again, we may
compensate the operator $\hat{T}$ by choosing $T_1$ sufficiently close to unity
or subsequently implementing the operator $\hat{T}^{-N^2}$, cf.
section~\ref{sec6.1}, so that we obtain
$\hat{Y}_\mathrm{tot}$ $\!\sim$ $\!\hat{F}$.
\section{\label{sec6}
         Examples of application}
We have seen that we may implement a given binomial in any moment of the
creation and annihilation operators of a number of traveling optical modes. A
repeated application then allows - in principle - the approximation of a desired
function of the respective mode operators. In a concrete example, the general
scheme can be simplified considerably. Let us illustrate this in some cases of
special interest such as single- and two-mode operators, since the application
of single-photon cloners may be avoided in this case.
\subsection{\label{sec6.1}
          Functions of the single-mode photon number operator}
Until now we have placed emphasis on the realization of general
operators. In a given case however, the respective scheme may be simplified
considerably. For example, let us assume that $\hat{A}$ $\!=$ $\!\hat{n}_0$.
This example is of particular importance since it allows the compensation of the
operators $\hat{T}$ in Eq.~(\ref{Y+-N}) and Eq.~(\ref{YgenN2}) for photon-number
truncated states as we will see below. In Eq.~(\ref{Yk}), we then have
$k$ $\!=$ $\!2$, $j_1$ $\!=$ $\!j_2$ $\!=$ $\!0$ and
$s_1$ $\!=$ $\!0$, $s_2$ $\!=$ $\!1$. Instead of applying the setup
Fig.~\ref{fig1} in the signal mode 0 first in case (0) and then in case (1), it
is sufficient to prepare its input ports 1 and 2 in a state $|\Psi\rangle_{12}$
as defined in Eq.~(\ref{s}). The resulting complete setup is shown in
Fig.~\ref{fig3}(a). If the photodetectors D$_1$ and D$_2$ detect 0 and 1
photons, respectively, the signal state $\hat{\varrho}$ is transformed in mode 0
according to Eq.~(\ref{Y}), where
\begin{eqnarray}
  \hat{Y}(\hat{A},\alpha,\beta)
  &=&\,_1\langle0|\,_2\langle1|\hat{U}_{21}\hat{U}_{01}|\Psi\rangle_{12}
  \nonumber\\
\label{Yn0}
  &=&\alpha T_1^{\hat{n}_0}(\hat{A}-\beta),
\end{eqnarray}
with
\begin{subequations}
\label{par1-a}
\begin{eqnarray}
  \hat{A}&=&\hat{n}_0,
  \\
  \alpha&=&-\frac{R_2|R_1|^2}{\sqrt{2}T_1},
  \\
  \beta&=&\left|\frac{T_1}{R_1}\right|^2
  \left(1+\frac{1}{T_1^*}\frac{T_2^{}}{R_2^{}}\right).
\end{eqnarray}
\end{subequations}
For fixed $\hat{U}_{01}$, $\beta$ can be arbitrarily chosen by varying the
parameters of $\hat{U}_{21}$. As a consequence, there is no need for cloning
devices. In order to realize a transformation of a photon-number truncated
signal state [cf. Eq.~(\ref{truncsig}) with $j_l$ $\!=$ $\!0$] according to
Eq.~(\ref{Y}) with a desired function $\hat{Y}$ $\!\sim$ $\!F(\hat{n}_0)$, there
have to be $N$ successive applications of Eq.~(\ref{Yn0}) with varying $\beta_n$
such that the overall operator becomes
\begin{eqnarray}
  \hat{Y}_\mathrm{tot}&=&
  \hat{Y}(\hat{A},\alpha_N,\beta_N)\cdots\hat{Y}(\hat{A},\alpha_1,\beta_1)
  \nonumber\\
  &=&\prod_{n=1}^N\left[
  \alpha_nT_1^{\hat{n}_0}(\hat{n}_0-\beta_n)\right]
  \nonumber\\
\label{Yn+-N}
  &=&\left(\prod_{n=1}^N\alpha_n\right)
  f_N^{-1}T_1^{N\hat{n}_0}F_N(\hat{n}_0).
\end{eqnarray}
Inserting $\hat{A}$ $\!=$ $\!\hat{n}_0$ and with it $A_l$ $\!=$ $\!l$ into
Eq.~(\ref{eigenpoly}), and furthermore substituting $T_1^{-Nl}F(l)$ for $F(l)$,
we see that the $\beta_n$ in Eq.~(\ref{Yn+-N}) must be chosen such that
\begin{equation}
\label{polyn0}
  F_N(\hat{n}_0)
  =\frac{1}{N!}\sum_{l=0}^N
  \binom{N}{l}
  (-T_1^N)^{-l}F(l)\prod_{\substack{k=0\\(k\neq l)}}^N(k-\hat{n}_0).
\end{equation}
Since then
$T_1^{N\hat{n}_0}F_N(\hat{n}_0)|n\rangle$ $\!=$ $\!F(\hat{n}_0)|n\rangle$
holds for number states $|n\rangle$ with $n$ $\!=$ $\!0,\ldots,N$, we obtain
\begin{equation}
\label{Ytotn0}
  \hat{Y}_\mathrm{tot}\hat{P}_N^{(0)}
  =\left(\prod_{n=1}^N\alpha_n\right)f_N^{-1}F(\hat{n}_0)\hat{P}_N^{(0)}
\end{equation}
[for $\hat{P}_N^{(0)}$ see Eq.~(\ref{truncsig})], where
\begin{equation}
  f_N=\frac{(-1)^N}{N!}\sum_{l=0}^N
  \binom{N}{l}
  (-T_1^N)^{-l}F(l),
\end{equation}
so that for photon-number truncated states the desired function $F(\hat{n}_0)$
is implemented.

As an illustration, let us consider the special case of the exponential
$F(\hat{n}_0)$ $\!=$ $\!z^{\hat{n}_0}$, which, if applied with appropriate $z$
in modes $j_1,$ $\!\ldots,$ $\!j_k$, allows the compensation of the operators
$\hat{T}$ in Eq.~(\ref{Y+-N}) and Eq.~(\ref{YgenN2}). Inserting
$F(l)$ $\!=$ $\!z^l$ into Eq.~(\ref{polyn0}) and applying
\begin{equation}
  \sum_{k=0}^n(-1)^k\binom{k}{l}\binom{x}{k}
  =\frac{n+1}{x-l}(-1)^n\binom{n}{l}\binom{x}{n+1},
\end{equation}
where $x$ is a complex number, $l$ $\!=$ $\!0,$ $\!1,\ldots$, and
\begin{equation}
  \binom{x}{k}
  \equiv\left\{\begin{array}{r@{\quad:\quad}l}
  1 & k=0 \\
  \frac{1}{k!}\prod_{j=0}^{k-1}(x-j) & k=1,2,\ldots,
  \end{array}\right.
\end{equation}
gives
\begin{eqnarray}
  F_N(\hat{n}_0)
  &=&\sum_{k=0}^N\left(T_1^{-N}z-1\right)^k\binom{\hat{n}_0}{k}
  \nonumber\\
\label{limitedpoly}
  &=&\sum_{k=0}^N\left(T_1^{-N}z-1\right)^k
  \frac{\hat{a}_0^{\dagger\,k}\hat{a}_0^k}{k!}.
\end{eqnarray}
Here, we have applied Eq.~(\ref{ordering2}) in the case of normal ordering,
$s$ $\!=$ $\!1$. Making use of Eq.~(\ref{ordering4}), we see that
\begin{eqnarray}
  F_N(\hat{n}_0)\hat{P}_N^{(0)}
  &=&\sum_{k=0}^\infty\left(T_1^{-N}z-1\right)^k
  \frac{\hat{a}_0^{\dagger\,k}\hat{a}_0^k}{k!}\hat{P}_N^{(0)}
  \nonumber\\
  &=&\;:\mathrm{e}^{\left(T_1^{-N}z-1\right)\hat{n}_0}:\;
  \hat{P}_N^{(0)}
  \nonumber\\
  &=&\left(T_1^{-N}z\right)^{\hat{n}_0}\hat{P}_N^{(0)},
\end{eqnarray}
where the symbol $:\;:$ denotes normal ordering, and Eq.~(\ref{Ytotn0}) becomes
\begin{equation}
  \hat{Y}_\mathrm{tot}\hat{P}_N^{(0)}
  =\left(\prod_{n=1}^N\alpha_n\right)
  N!\left(T_1^{-N}z-1\right)^{-N}z^{\hat{n}_0}\hat{P}_N^{(0)}.
\end{equation}
\subsection{\label{sec6.2}
            Cross-Kerr nonlinearity}
\begin{figure}[ht]
\includegraphics[width=8.6cm]{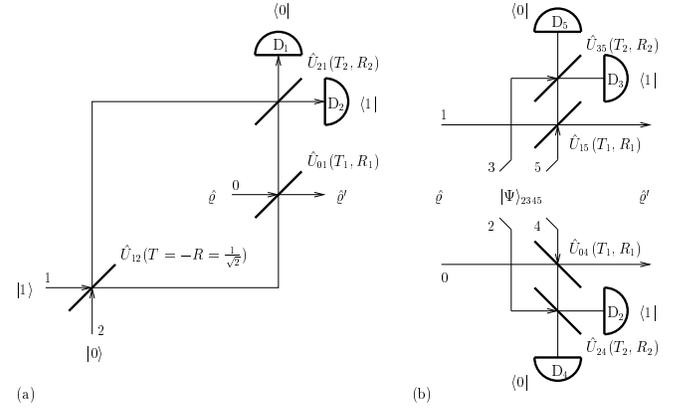}
\caption{\label{fig3}
Setups consisting of beam splitters $\hat{U}$ and photodetectors D,
conditionally implementing a desired binomial in $\hat{n}_0$ (a) and a desired
binomial in $(\hat{n}_1-\gamma)(\hat{n}_0-\gamma)$ with some given $\gamma$ (b).
}
\end{figure}

Another example allowing a simplification of the general configuration is the
implementation of a two-mode cross-Kerr interaction,
$\hat{Y}$ $\!\sim$ $\!\mathrm{e}^{\mathrm{i}\varphi\hat{n}_1\hat{n}_0}$. 
In Eq.~(\ref{Yk}), we then have
$\hat{A}$ $\!=$ $\!\hat{n}_1\hat{n}_0$,
$k$ $\!=$ $\!4$,
$j_1$ $\!=$ $\!j_2$ $\!=$ $\!0$,
$j_3$ $\!=$ $\!j_4$ $\!=$ $\!1$, and
$s_1$ $\!=$ $\!s_3$ $\!=$ $\!0$,
$s_2$ $\!=$ $\!s_4$ $\!=$ $\!1$.
Instead of applying the setup Fig.~\ref{fig1} in the signal modes 0 and 1 first
in case (0) and then in case (1), a single application in each signal mode as
shown in Fig.~\ref{fig3}(b) is sufficient. To the setup acting on mode 1
consisting of beam splitters $\hat{U}_{15}(T_1,R_1)$ and $\hat{U}_{35}(T_2,R_2)$
as well as photodetectors D$_3$ and D$_5$, a second, identical setup acting on
mode 0 is added consisting of beam splitters $\hat{U}_{04}(T_1,R_1)$ and
$\hat{U}_{24}(T_2,R_2)$ as well as photodetectors D$_2$ and D$_4$. The modes are
labeled as shown in Fig.~\ref{fig3}(b).
The input modes are prepared in a state $|\Psi\rangle_{2345}$ as given by
Eq.~(\ref{Psi1234}) with modes relabeled accordingly. If the photodetectors
D$_2$ and D$_3$ detect 1 whereas D$_4$ and D$_5$ detect 0 photons, respectively,
the signal state $\hat{\varrho}$ is transformed in modes 0 and 1 according to
Eq.~(\ref{Y}), where
\begin{eqnarray}
  \hat{Y}(\hat{A},\alpha,\beta)
  &=&\,_{23}\langle1|\,_{45}\langle0|
  \hat{U}_{24}\hat{U}_{04}\hat{U}_{35}\hat{U}_{15}
  |\Psi\rangle_{2345}
  \nonumber\\
\label{Yn1n0}
  &=&\alpha T_1^{\hat{n}_1+\hat{n}_0}\bigl(\hat{A}-\beta\bigr),
\end{eqnarray}
with
\begin{subequations}
\begin{eqnarray}
  \hat{A}&=&(\hat{n}_1-\gamma)(\hat{n}_0-\gamma),
  \\
  \alpha&=&-\frac{T_2^2}{\beta\sqrt{1+|\xi|^2}},
  \\
  \beta&=&-\frac{1}{\xi}\left(\frac{T_1}{|R_1|^2}\right)^2
  \left(\frac{T_2}{R_2}\right)^2,
  \\
  \gamma&=&\left|\frac{T_1}{R_1}\right|^2.
\end{eqnarray}
\end{subequations}
For fixed $\xi$, $T_1$ and $R_1$, $\beta$ can be chosen as desired by varying 
$T_2$ and $R_2$. By $N$ repeated applications of Eq.~(\ref{Yn1n0}) with varying
$\beta_n$ we obtain
\begin{eqnarray}
  \hat{Y}_\mathrm{tot}
  &=&\hat{Y}(\hat{A},\alpha_N,\beta_N)\cdots\hat{Y}(\hat{A},\alpha_1,\beta_1)
  \nonumber\\
\label{Yn1n0N}
  &=&\left(\prod_{n=1}^N\alpha_n\right)
  f_N^{-1}T_1^{N(\hat{n}_1+\hat{n}_0)}F_N.
\end{eqnarray}
We now tune $T_1$ and $R_1$ such that $\gamma$ $\!=$ $\!\varphi^{-1}2k\pi$
with some arbitrary $k$ $\!=$ $\!1,2,\cdots$. The $\beta_n$ are chosen such
that in Eq.~(\ref{Yn1n0N}) the polynomial $F_N$ of order $N$ in $\hat{A}$
becomes
\begin{equation}
  F_N=\sum_{l=0}^N\mathrm{e}^{\mathrm{i}\varphi A_l}
  \prod_{\substack{k=0\\(k\neq l)}}^N\frac{\hat{A}-A_k}{A_l-A_k},
\end{equation}
where $A_0,\ldots,A_N$ are the ($N$ $\!+$ $\!1$) distinct values of the
expression $(n_1-\gamma)(n_0-\gamma)$ with $n_{0,1}$ $\!=$ $\!0,\ldots,M$
[cf. the remarks following Eq.~(\ref{eigenstates})].
On signal states $\hat{\varrho}$ whose photon number is limited to some given
value $M$ in mode 0 and 1, i.e., $\hat{\varrho}$ satisfies Eq.~(\ref{truncsig}),
where $\hat{P}_N^{(j_l)}$ is replaced with
$\hat{P}_M^{(0)}\otimes\hat{P}_M^{(1)}$, Eq.~(\ref{Yn1n0N}) then acts like an
operator
\begin{equation}
\label{kerrtot}
  \hat{Y}_\mathrm{tot}
  =\left(\prod_{n=1}^N\alpha_n\right)f_N^{-1}T_1^{N(\hat{n}_1+\hat{n}_0)}
  \mathrm{e}^{\mathrm{i}\varphi\hat{n}_1\hat{n}_0}.
\end{equation}
In order to implement the unitary operator
$\mathrm{e}^{\mathrm{i}\varphi\hat{n}_1\hat{n}_0}$, $T_1$ may be chosen
sufficiently close to unity. Again, there is the alternative of a precise
compensation of the extra exponential in Eq.~(\ref{kerrtot}) by an additional
implementation of the operator $T_1^{-N(\hat{n}_1+\hat{n}_0)}$ as discussed in
the previous section~\ref{sec6.1}. In this way, we have proposed a beam splitter
arrangement relying on zero- and single-photon preparations and detections that
acts on photon-number truncated states like a two-mode Kerr nonlinearity.
\subsection{\label{sec6.3}
            U($N$)}
Other operators whose action on photon-number truncated signal states can be
realized exactly by a polynomial are those performing a U($N$) transformation.
Since they can be factorized into U(2) couplers $\hat{U}_{jk}$ of the respective
modes, let us limit attention to the U(2) beam splitter operator $\hat{U}_{jk}$
defined in the beginning by Eq.~(\ref{MC}), which can equivalently be written in
the form of \cite{clausen6},
\begin{equation}
\label{UC}
  \hat{U}_{jk}=
  (\mathcal{P}\mathcal{T})^{\hat{n}_j}
  \mathrm{e}^{-\mathcal{P}\mathcal{R}^*\hat{a}_k^\dagger\hat{a}_j^{}}
  \mathrm{e}^{\mathcal{P}^*\mathcal{R}\hat{a}_j^\dagger\hat{a}_k^{}}
  (\mathcal{P}^*\mathcal{T})^{-\hat{n}_k}.
\end{equation}
Its parameters are here written in calligraphic style to distinguish from those
of the beam splitters in the device. The implementation of exponentials of
$\hat{a}_k^\dagger\hat{a}_j^{}$ with $k$ $\!=$ $\!j$ explained in
section~\ref{sec6.1} represents the case of section~\ref{sec4.1} and those with
$k$ $\!\neq$ $\!j$ represents the case of section~\ref{sec4.2}. Together with a
detection of the vacuum state, the latter can be used to achieve a teleportation
of a state $\hat{\varrho}$ from mode $j$ to mode $k$,
\begin{equation}
  \hat{\varrho}_k=\hat{I}_{kj}\hat{\varrho}_j\hat{I}_{jk},
\end{equation}
where
\begin{equation}
\label{Ikj}
  \hat{I}_{kj}=\sum_{n=0}^{\infty}|n\rangle_k\,_j\langle n|
  =\,_j\langle0|\mathrm{e}^{\hat{a}_k^\dagger\hat{a}_j^{}}|0\rangle_k
  =\hat{I}_{jk}^\dagger.
\end{equation}
A comparison with Eq.~(\ref{UC}) shows that
$\exp(\hat{a}_k^\dagger\hat{a}_j^{})$ [or $\exp(\hat{a}_j^\dagger\hat{a}_k^{})$]
is realized for $\mathcal{T}$ $\!=$ $\!0$. Physically, the beam splitter then
describes a mirror. Since Eq.~(\ref{UC}) becomes singular in this case, we have
to construct it by a successive implementation of two operators Eq.~(\ref{UC})
whose combined action results according to
\begin{eqnarray}
  &&\hat{U}_{jk}(\mathcal{T}_2,\mathcal{R}_2,
  \mathcal{P}_2)\hat{U}_{jk}(\mathcal{T}_1,\mathcal{R}_1,\mathcal{P}_1)
  \nonumber\\&=&
  \hat{U}_{jk}(\mathcal{T}_2^{}\mathcal{T}_1^{}-\mathcal{R}_2^{}\mathcal{R}_1^*,
  \mathcal{T}_2^{}\mathcal{R}_1^{}+\mathcal{R}_2^{}\mathcal{T}_1^*,
  \mathcal{P}_2^{}\mathcal{P}_1^{})
\end{eqnarray}
again in a U(2) beam splitter operator. Choosing
$\mathcal{T}_j$ $\!=$ $\!\mathcal{R}_j$ $\!=$ $\!1/\sqrt{2}$ and
$\mathcal{P}_j$ $\!=$ $\!1$, we get an operator 
$\hat{U}_{jk}(0,1,1)$. If implemented by our scheme with separated modes $j$ and
$k$, such a \textquoteleft{quantum mirror}\textquoteright\, realizes a
bidirectional teleportation between these modes.

More generally, we may consider a transformation in $N$ modes
\mbox{$j$ $\!=$ $\!1,\ldots,N$} of a signal according to (\ref{Y}) by an
operator $\hat{Y}$ $\!\sim$ $\!\hat{U}$ whose action is defined by
\begin{equation}
  \hat{U}^\dagger\hat{a}_j\hat{U}=\hat{a}_{p(j)},
\end{equation}
where \mbox{$p(1),\ldots,p(N)$} is a permutation of \mbox{$1,\ldots,N$}. The set
of such permutations of $N$ mode indices forms the discrete symmetric subgroup
S($N$) of U($N$) physically representing $N$-mode mirrors (with the identical
operation included). Note that in general we may add additional phase shifts
according to $\hat{U}^\prime$ $\!=$
$\!\mathrm{e}^{\mathrm{i}\sum_{j=1}^N\varphi_j\hat{n}_j}\hat{U}$, so that S($N$)
is replaced with U(1)$^N\,\otimes\,$S($N$). If $\hat{U}$ is implemented by the
setup Fig.~\ref{fig2} with spatially separated signal modes, then we again have
the situation of a teleportation. One possibility to construct $\hat{U}$ is to
introduce auxiliary modes \mbox{$-1,\ldots,-N$}, and to apply (\ref{Ikj})
repeatedly according to
\begin{equation}
  \hat{U}=\hat{I}_{p(N),-N}\cdots\hat{I}_{p(1),-1}
  \hat{I}_{-N,N}\cdots\hat{I}_{-1,1}.
\end{equation}
\subsection{\label{sec6.4}
            Preparation of multi-photon entangled states}
We conclude with an example of state preparation which may be understood as a
state transformation with a given input state. Consider the following situation.
By feeding the overall setup with vacuum,
$\hat{\varrho}$ $\!=$ $\!|0\rangle\langle0|$, the scheme may be used to generate
a state difficult to prepare by other means. An example are $k$-mode states
\begin{eqnarray}
\label{multisqueezed}
  |z\rangle_{1,\ldots,k}
  &=&\sqrt{\frac{1-|z|^2}{1-|z|^{2(N+1)}}}\sum_{n=0}^N z^n|n\rangle_{1,\ldots,k}
  \\
  &=&\frac{1}{\sqrt{p}}\hat{Y}|0\rangle_{1,\ldots,k},
  \nonumber
\end{eqnarray}
which can be generated by applying a polynomial
\begin{equation}
  \hat{Y}\sim\sum_{n=0}^Nn!^{-\frac{k}{2}}(z\hat{A})^n
\end{equation}
in $\hat{A}$ $\!=$ $\!\hat{a}_k^\dagger\cdots\hat{a}_1^\dagger$ to the vacuum.
The states Eq.~(\ref{multisqueezed}) are of interest from the theoretical point
of view. Consider the case $|z|$ $\!\le$ $\!1$ and $N$ $\!\to$ $\!\infty$.
We see that for $k$ $\!=$ $\!1$ we obtain a coherent phase state (in particular,
for $|z|$ $\!=$ $\!1$ a London phase state). For $k$ $\!=$ $\!2$ we obtain a
two-mode squeezed state (in particular, for $|z|$ $\!=$ $\!1$ an EPR-like
state). In the limit $k$ $\!\to$ $\!\infty$, the states
Eq.~(\ref{multisqueezed}) may be used as a representation of an EPR-like state
between a number $n$ of quantum \textit{fields} (instead of two modes of a
field). To see this, the fields are indexed according to
$\lceil{j}\rceil$ $\!\equiv$ $\!1$ $\!+$
$\![(j$ $\!-$ $\!1)$ $\!\mathrm{mod}$ $\!n]$, i.e., mode $j$ is ascribed to
field $\lceil{j}\rceil$.
\section{\label{sec7}
         Conclusion}
In summary it can be said that, starting from a single photon, cf.
Eq.~(\ref{s}), we can with a given probability approach arbitrary operators of
spatially separated traveling optical modes. The problem is that the
success probability, which depends on the desired transformation and the signal
state in general, is expected to decline exponentially with an increasing number
of steps $N$, so that the practical applicability of the scheme is limited to
small $N$, sufficient to engineer, e.g., states in the vicinity of the
vacuum. On the other hand, the application of giant nonlinearities to fields
containing only few photons represents just the case discussed in potential
quantum information processing devices. To give an estimation, consider a
passively mode-locked laser whose resonator has an optical round trip length of
$l$ = 3 cm, so that the repetition frequency of the emitted pulse train is
$\nu$ = $l^{-1}c$ = 10 GHz. If this pulse train is used to prepare the entangled
state needed to run the scheme, then its repetition frequency determines that of
the overall device. An assumed total success probability of
$p$ $\!=$ $\!10^{-10}$ would then reduce the mean repetition frequency of the
properly transformed output pulses to $\nu^\prime$ = 1 Hz, which may still be
useful for basic research experiments but unacceptable for technical
applications. Apart from this, there may be the situation where only a single
unknown signal pulse is available that must not be spoiled by a wrong detection
since no copies can be produced of it. This constitutes the main drawback of all
conditional measurement schemes.

It may be worth mentioning that the whole scheme may also run 
\textquoteleft{time-reversed}\textquoteright\,, i.e., all pulses are sent in
opposite directions through the device, while the locations of state
preparations and (assumed) detections are interchanged, so that the resulting
\textquoteleft{adjoint}\textquoteright\, scheme implements the adjoint operator
$\hat{Y}^\dagger$.
\begin{acknowledgments}
This work was supported by the Deutsche Forschungsgemeinschaft.
\end{acknowledgments}
\appendix
\section{\label{secA1}
         Single-photon cloning}
In what follows, we give a number of possibilities to implement a single-photon
cloner Eq.~(\ref{Q}).
\subsection{\label{secA1.1}
            Cross-Kerr nonlinearity}
To implement single-photon cloning, we may apply a Mach-Zehnder-interferometer
equipped with a cross-Kerr nonlinearity as can be seen as follows. We prepare
input mode 2 of a balanced beam splitter
$\hat{U}_{21}(T$ $\!=$ $\!-R$ $\!=$ $\!1/\sqrt{2})$ in a single-photon state
$|1\rangle$ and the other input mode in the vacuum state $|0\rangle$. After
output mode 2 has passed a cross-Kerr coupler
\begin{equation}
  \hat{U}_{02}=\mathrm{e}^{\mathrm{i}\pi\hat{n}_2\hat{n}_0},
\end{equation}
it is remixed with output mode 1 at a second beam splitter
$\hat{U}_{21}^\dagger$. The output mode 2 of this second beam splitter is then
again mixed with an auxiliary mode 3 prepared in a coherent state $|\xi\rangle$
with $\xi$ $\!=$ $\!T^{-1}R$ using a beam splitter $\hat{U}_{32}(T,R)$. If
eventually 0 and 1 photons are detected in modes 2 and 3, respectively, the
action of the complete setup on the state of the input mode 0 can be described
by an operator
\begin{eqnarray}
  \hat{Y}&=&\,_2\langle0|\,_3\langle1|
  \hat{U}_{32}\hat{U}_{21}^\dagger\hat{U}_{02}\hat{U}_{21}
  |0\rangle_1|1\rangle_2|\xi\rangle_3
  \nonumber\\
  &=&\mathrm{e}^{\mathrm{i}\arg{R}}\sqrt{p}\,
  \sum_{s=0}^1\frac{1+(-1)^{\hat{n}_0+s}}{2}|s\rangle_1,
\end{eqnarray}
so that
\begin{equation}
  \hat{Y}\hat{P}_1^{(0)}=\mathrm{e}^{\mathrm{i}\arg{R}}\sqrt{p}\;\hat{Q}_{10}
\end{equation}
[for $\hat{P}_1^{(0)}$ see Eq.~(\ref{truncsig})]. The success probability
\begin{equation}
  p=\left[(1+|\xi|^{-2})\mathrm{e}^{|\xi|^2}\right]^{-1}
\end{equation}
attains for $|\xi|^2$ $\!=$ $\!(\sqrt{5}-1)/2$ $\!\approx$ $\!0.62$ its maximum
value of $p_\mathrm{max}$ $\!\approx$ $\!0.21$.
\subsection{\label{secA1.2}
            Three-wave-mixer}
A disadvantage of the above proposal is its dependency on a given measurement
result. Unitary photon cloning can be achieved using a three-wave-mixer
\begin{equation}
  \hat{U}_{012}(\varphi)
  =\mathrm{e}^{\mathrm{i}\varphi(\hat{a}_0^{}\hat{a}_1^\dagger\hat{a}_2^\dagger
  +\hat{a}_0^\dagger\hat{a}_1^{}\hat{a}_2^{})}.
\end{equation}
Preparing its input mode 0 in a photon number state $|s\rangle$ with
$s$ $\!=$ $\!0,1$ and its input modes 1 and 2 in the vacuum state $|0\rangle$,
the output state becomes
\begin{equation}
  \hat{U}_{012}(\varphi)|s\rangle_0|0\rangle_{12}
  =\cos(s\varphi)|s\rangle_0|0\rangle_{12}
  +\mathrm{i}\sin(s\varphi)|0\rangle_0|s\rangle_{12},
\end{equation}
so that for a phase of $\varphi$ $\!=$ $\!\pi/2$ and an additional application
of a preceding phase shifter $\mathrm{e}^{-\mathrm{i}\frac{\pi}{2}\hat{n}_0}$ we
obtain an output state $\hat{Y}|s\rangle_0$ $\!=$ $\!|s\rangle_{12}$, where
\begin{equation}
  \hat{Y}=\,_0\langle0|
  \hat{U}_{012}(\pi/2)\mathrm{e}^{-\mathrm{i}\frac{\pi}{2}\hat{n}_0}
  |0\rangle_{12}.
\end{equation}
After relabeling output mode 2 as 0 we get
\begin{equation}
  \hat{Y}\hat{P}_1^{(0)}=\hat{Q}_{10},
\end{equation}
cf. Eq.~(\ref{Q}) and Eq.~(\ref{truncsig}). Formally, we could consider a
success probability for which
$p$ $\!=$ $\!\,_0\langle s|\hat{Y}^\dagger\hat{Y}|s\rangle_0$ $\!=$ $\!1$ holds
as expected for an unconditional unitary operation.
\subsection{\label{secA1.3}
            Linear optics}
A disadvantage of the previous two proposals is their dependency on large
nonlinear coefficients which are hardly achievable in practice. Therefore, we
now give a possibility of implementing Eq.~(\ref{Q}) solely based on beam
splitters and zero- and single-photon detections. Assume that a state
\begin{equation}
\label{Psi1234}
  |\Psi\rangle_{1234}
  =\frac{|1\rangle_{12}|0\rangle_{34}+\xi\,|0\rangle_{12}|1\rangle_{34}}
  {\sqrt{1+|\xi|^2}}
\end{equation}
with some arbitrary $\xi$ $\!\neq$ $\!0$ is prepared. Mode 3 is mixed with an
auxiliary mode 5 prepared in a coherent state $|\xi\rangle$ using a beam
splitter $\hat{U}_{53}(T$ $\!=$ $\!R$ $\!=$ $\!1/\sqrt{2})$ and mode 4 passes
a beam splitter $\hat{U}_{04}(T$ $\!=$ $\!R$ $\!=$ $\!1/\sqrt{2})$. If
eventually 0 photons are detected in modes 3 and 4 but 1 photon in each of the
modes 0 and 5, respectively, the device realizes an operator
\begin{eqnarray}
  \hat{Y}&=&\,_{34}\langle0|\,_{05}\langle1|
  \hat{U}_{04}\hat{U}_{53}|\Psi\rangle_{1234}|\xi\rangle_5
  \nonumber\\
\label{YQ}
  &=&\mathrm{e}^{\mathrm{i}\arg{\xi}}\sqrt{p}\,\hat{Q}_{10}
\end{eqnarray}
that acts on states $|s\rangle$ in which input port 0 of the beam splitter
$\hat{U}_{04}$ is prepared. Note that in the second line of Eq.~(\ref{YQ}) 
we have relabeled output mode 2 as mode 0. The success probability
\begin{equation}
  p=\left[4(1+|\xi|^{-2})\mathrm{e}^{|\xi|^2}\right]^{-1}
\end{equation}
becomes maximum for
$|\xi|^2$ $\!=$ $\!(\sqrt{5}-1)/2$ $\!\approx$ $\!0.62$ for which 
$p$ $\!\approx$ $\!0.05$.

\begin{figure}[ht]
\includegraphics[width=8.6cm]{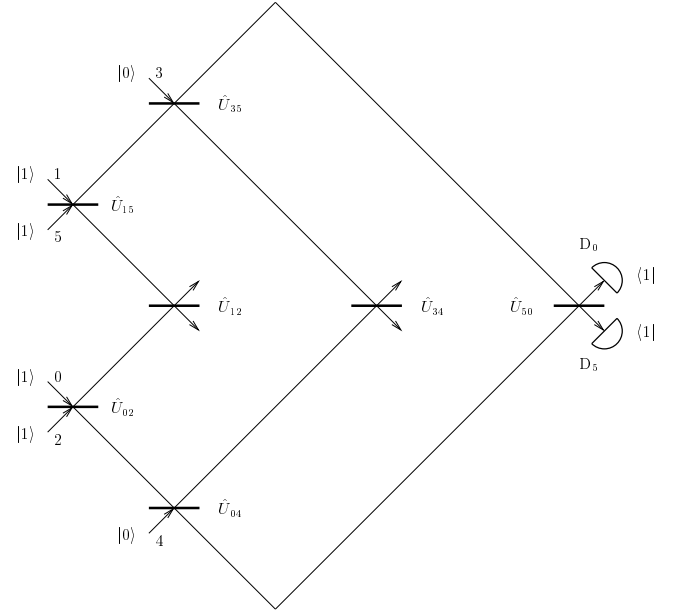}
\caption{\label{fig4}
Setup consisting of beam splitters $\hat{U}$ and photodetectors D, conditionally
preparing a state given by Eq.~(\ref{Psi1234}).
}
\end{figure}

One way to prepare the state Eq.~(\ref{Psi1234}) from single-photon states is
shown in Fig.~\ref{fig4}. If transmittance and reflectance of the beam splitter
$\hat{U}_{04}$ are denoted by $T$ and $R$, respectively, then those of beam
splitter $\hat{U}_{35}$ are given by $T^*$ and $R^*$. The parameters of all
other beam splitters are given by $T$ $\!=$ $\!R$ $\!=$ $\!1/\sqrt{2}$.
The setup represents an array of these beam splitters which is fed with a number
of single-photon states $|1\rangle$ and vacuum states $|0\rangle$ as shown in
the figure. If each of the two photodetectors detects 1 photon, the output state
becomes
\begin{equation}
  \frac{\mathrm{e}^{\mathrm{i}\varphi}}{\sqrt{p}}\,_{05}\langle1|
  \hat{U}_{50}\hat{U}_{34}\hat{U}_{04}\hat{U}_{35}
  \hat{U}_{12}\hat{U}_{02}\hat{U}_{15}
  |1\rangle_{0125}|0\rangle_{34}=|\Psi\rangle_{1234},
\end{equation}
where $\varphi$ $\!=$ $\!\pi$ $\!-$ $\!2\arg{T}$ is an unrelevant phase and
\begin{equation}
  \xi=-R^{*2}.
\end{equation}
The success probability reads
\begin{equation}
  p=\left(\frac{1-|\xi|}{2}\right)^2(1+|\xi|^2).
\end{equation}
We see that $p$ drops from its maximum value of 0.25 as attained for
$\xi$ $\!=$ $\!0$ to 0 if $|\xi|$ $\!\to$ $\!1$. For the above value
of $|\xi|^2$ $\!=$ $\!0.62$ we obtain $p$ $\!\approx$ $\!0.02$.
\section{\label{secA2}
         $s$-ordering relations for the photon number operator}
In what follows, we compile some relations regarding the $s$-ordering of 
the photon number operator $\hat{n}$ $\!=$ $\!\hat{a}^\dagger\hat{a}$ as used in
section~\ref{sec6.1}. We start with its $k$th power. Introducing $s$-ordering
according to \cite{orderingGlauber},
\begin{equation}
\label{ordering1}
  \{\hat{a}^{\dagger\,m}\hat{a}^n\}_s
  =\!\!\!\sum_{k=0}^{\min[m,n]}
  k!\binom{m}{k}\!\binom{n}{k}\!\left(\frac{t\!-\!s}{2}\right)^k
  \{\hat{a}^{\dagger\,m-k}\hat{a}^{n-k}\}_t,
\end{equation}
and applying \cite{Prudnikov1}, p. 626, no. 30, we get
\begin{eqnarray}
  \left\{\hat{n}^k\right\}_s
  &=&k!\sum_{l=0}^k\left(\frac{1\!-\!s}{2}\right)^{k-l}
  \binom{k}{l}\binom{\hat{n}}{l}
  \nonumber\\
\label{ordering2}
  &=&k!\left(-\frac{s\!+\!1}{2}\right)^k\mathrm{P}_k^{(0,\hat{n}-k)}
  \!\left(\frac{s\!-\!3}{s\!+\!1}\right).
\end{eqnarray}
The inverse relation is obtained from Eq.~(\ref{ordering1}) with $s$ $\!=$ $\!1$
and $t$ renamed as $s$, inserting Eq.~(\ref{ordering2}) with $s$ $\!=$ $\!1$,
and applying \cite{Prudnikov1}, p. 624, no. 25, as well as p. 614, no. 31, which
yields
\begin{eqnarray}
  \hat{n}^k
  \!&=&\!\left\{\sum_{m,j,l=0}^k\binom{j}{m}\!\binom{j}{l}\!(-1)^{m+j}m^k\!
  \left(\frac{s\!-\!1}{2}\right)^{j-l}\frac{\hat{n}^l}{l!}\right\}_s\;\;\;\;
  \nonumber\\
\label{ordering3}
  \!&=&\!\left\{\sum_{m,j=0}^k\binom{j}{m}(-1)^mm^k
  \left(\frac{1\!-\!s}{2}\right)^j\mathrm{L}_j
  \!\left(\frac{2\hat{n}}{1\!-\!s}\right)\right\}_s\!\!\!\!.
\end{eqnarray}
To consider the case where $\hat{n}$ is the exponent, we apply
\cite{Prudnikov1}, p. 709, no. 3, as well as p. 612, no. 1, to
Eq.~(\ref{ordering2}), which gives
\begin{equation}
\label{ordering4}
  \left\{\mathrm{e}^{\alpha\hat{n}}\right\}_s
  =\frac{\left(1+\frac{s+1}{2}\alpha\right)^{\hat{n}}}
  {\left(1+\frac{s-1}{2}\alpha\right)^{\hat{n}+1}}.
\end{equation}
The inverse relation
\begin{equation}
\label{ordering5}
  \alpha^{\hat{n}}
  =\left\{\frac{\mathrm{e}^{\left(\frac{1}{\alpha-1}+
  \frac{1-s}{2}\right)^{-1}\hat{n}}}{1+\frac{(1-s)(\alpha-1)}{2}}\right\}_s
\end{equation}
then follows directly from Eq.~(\ref{ordering4}) in accordance with
\cite{orderingGlauber}.

\end{document}